\documentclass[hyper,letterpaper,12pt]{article}
\usepackage{a4wide}
\usepackage{amsmath}
\usepackage[
      colorlinks=true,
      linkcolor=blue,
      urlcolor=blue,
      filecolor=black,
      citecolor=blue,
      pdfstartview=FitV,
        pdfpagemode=None,
        bookmarksopen=true
      ]{hyperref}
\usepackage{color}
\usepackage{graphicx}
\usepackage{amsfonts}
\usepackage{amssymb}

\def\bc{\begin{center}}
\def\nno{\nonumber}
\def\ec{\end{center}}
\def\be{\begin{eqnarray}}
\def\ee{\end{eqnarray}}

\definecolor{dyellow}{rgb}{1.,0.8,.0}
\definecolor{myblue}{rgb}{.1,.1,.7}
\definecolor{dcyan}{rgb}{.0,.6,.6}
\definecolor{dmagenta}{rgb}{0.6,0.0,0.6}
\definecolor{brown}{rgb}{0.6,0.2,0.}
\definecolor{darkblue}{rgb}{.0,.0,0.5}
\definecolor{darkred}{rgb}{0.75,0.0,0.0}
\definecolor{orange}{rgb}{1.,.6,.0}
\definecolor{dorange}{rgb}{0.8,.4,.0}
\definecolor{darkgreen}{rgb}{0.0,0.6,0.0}
\definecolor{purple}{rgb}{.4,.0,.4}
\definecolor{lightgrey}{rgb}{0.7, 0.7, 0.7}
\definecolor{grey}{rgb}{0.4, 0.4, 0.4}
\def\la{\lambda}

\def\om{\omega}

\def\pa{\partial}

\def\Dl{\Delta}

\def\d#1#2{\frac{\displaystyle #1}{\displaystyle #2}}

\newcommand\rha{\rightarrow}

\def\d{{\rm d}}

\begin{document}

\title{\bf Magnetic Field Effect on the Phase Transition in AdS Soliton  Spacetime}

\author{\large
~Rong-Gen Cai\footnote{E-mail: cairg@itp.ac.cn}~, ~~Li
Li\footnote{E-mail: liliphy@itp.ac.cn}~,~~Hai-Qing
Zhang\footnote{E-mail: hqzhang@itp.ac.cn}~,
~~Yun-Long Zhang\footnote{E-mail: zhangyl@itp.ac.cn}\\
\\
\small State Key Laboratory of Theoretical Physics,\\
\small Institute of Theoretical Physics, Chinese Academy of Sciences,\\
\small P.O. Box 2735, Beijing 100190, People's Republic of China\\}

\date{\small (\today)}

\maketitle

\begin{abstract}
\normalsize

We investigate the scalar perturbations in an anti-de Sitter soliton background
coupled to a Maxwell field via marginally stable modes. In the probe
limit, we study the magnetic field effect on the holographic
insulator/superconductor phase transition numerically and
analytically. The condensate will be localized in a finite circular
region for any finite constant magnetic field.  Near the critical
point, we find that there exists a simple relation among the
critical chemical potential, magnetic field, the charge and mass of
the scalar field. This relation indicates that the presence of the
magnetic field causes the transition from insulator to superconductor to be difficult.
\end{abstract}

\section{ Introduction}

The AdS/CFT correspondence~\cite{Maldacena:1997re} provides a
powerful theoretical method to understand the strongly coupled field
theories in condensed matter physics. A holographic superconductor
(superfluid) model has been constructed recently in the work
\cite{Gubser:2008px,Hartnoll:2008vx}(for reviews see
\cite{Hartnoll:2009sz}). The dual gravitational configurations are
some anti-de Sitter (AdS) black holes with/without some charged matter contents.

The phase transition studied in Refs.\cite{Gubser:2008px,Hartnoll:2008vx}
is actually a holographic superconductor/metal phase transition. The
holographic superconductor model can be simply constructed by an
Einstein-Maxwell theory with a negative cosmological constant
coupled to a complex scalar field. In particular, when the
temperature of the black hole is below a critical temperature, there
are at least two distinct mechanisms leading to the black hole
solution unstable to develop a scalar hair near the horizon
\cite{Hartnoll:2008kx}. And the condensation of the scalar hair
induces the local U(1) symmetry breaking of the system, which gives
the nonvanishing vacuum expectation value to the dual charged
operator in the boundary field theory. Therefor, the U(1) symmetry
breaking in gravity leads to a breaking of the global U(1) symmetry
in the dual field theory. This results in a superconductor
(superfluid) phase transition.

The holographic insulator/superconductor phase transition  was first
researched in Ref.\cite{Nishioka:2009zj}. A five-dimensional AdS soliton
background \cite{Horowitz:1998ha} coupled to a Maxwell field and a
scalar field was used to model the holographic
insulator/superconductor phase transition at zero temperature. The
normal phase in the AdS soliton is dual to a confined gauge theory
with a mass gap which is reminiscent of an insulator phase
\cite{Witten:1998zw}. As the chemical potential grows sufficiently
up to a critical value, the instability is triggered, resulting in
the emergence of the scalar hair which is dual to a superconducting
phase in the boundary field theory. The holographic
insulator/superconductor phase transition was also investigated in
Refs.\cite{Horowitz:2010jq,Akhavan:2010bf,Basu:2011yg,Brihaye:2011vk,
Cai:2011ky,Peng:2011gh,Pan:2011ah}.

There are some discussions of holographic superconductors in the
presence of magnetic field
\cite{Hartnoll:2008kx,Nakano:2008xc,Albash:2008eh,Wen:2008pb,Maeda:2008ir,Ge:2010aa,arXiv:0903.1864},
but they mainly focused on superconductor/metal phase transition.
Motivated by these studies and especially by our previous works
\cite{Cai:2011ky,Cai:2011qm}, we expect to explore how the magnetic
field impacts the behavior of the condensate if a magnetic field
is added into the AdS soliton spacetime. We are working in the probe
limit, which means that the scalar field and Maxwell field have no
back-reaction to the gravity background.

In this paper, we first adopt the idea of marginally stable modes
\cite{Gubser:2008px} to study the holographic
insulator/superconductor phase transition in AdS soliton background
at zero temperature. We employ the method introduced in
Ref.\cite{Horowitz:1999jd} to study the quasinormal modes(QNMs) of the
scalar perturbations of the system just like the one in
Ref.\cite{Nishioka:2009zj}. At some critical values of the chemical
potential $\mu$ and magnetic field $B$, the marginally stable modes
will emerge. This represents that at the critical parameters, the AdS
soliton background becomes unstable and will be fond of an AdS
soliton background coupled with the hair of charged scalar fields.
In particular, for some given mass $m^2$ and charge $q$, the special
combination of the critical chemical potential $\mu$ and the
critical magnetic field $B$ satisfies a simple relation, {\it i.e.}
$q^2\mu^2-|qB|=\Lambda^2$, where $\Lambda$ is a positive constant.
The minus sign in the combination is interesting because, as the
magnetic field grows stronger, the critical chemical potential
becomes higher. This indicates that the presence of the magnetic
field causes the transition from insulator to superconductor to be difficult.
Setting the magnetic field
vanishing, the critical chemical potentials we derived are
consistent with the results obtained by previous works
\cite{Cai:2011ky,Cai:2011qm}. Actually, corresponding to various
critical parameters, there are multiple marginally stable modes
related to the nodes $n=1, 2, 3 \cdots$, which are unstable due to
the oscillations of the field in the radial direction
\cite{Gubser:2008px, Gubser:2008wv}. Taking advantage of the
shooting method, we plot the profile of the scalar field depending
on the radial direction. From these diagrams, one can intuitively
see the ``nodes" of the scalar field. Furthermore, using the
variational method for the Sturm-Liouville eigenvalue problem
\cite{Siopsis:2010uq}, we analytically study the holographic
insulator/superconductors phase transition, following the previous
work in Ref.\cite{Cai:2011ky}. Near the critical point, we show a simple
relation among  the critical chemical potential, magnetic field, the
charge and mass of the scalar field.

The paper is organized as follows: in Sec.\ref{sect:back}, we
introduce the AdS soliton background and obtain the equations of
motion in the probe limit. We study the marginally stable modes in
Sec.\ref{sect:QNMs}. In Sec.\ref{sect:shoot} the system is
solved by shooting method. In Sec.\ref{sect:sturm} we extract
the relation of the four parameters at the critical phase transition
points using the variational method for the Sturm-Liouville
eigenvalue problem. Conclusions and discussions are drawn in
Sec.\ref{sect:con}.

\section{The Background}
\label{sect:back}

We construct the model of the holographic insulator/superconductor
phase transition with the Einstein-Maxwell-scalar action in
five-dimensional spacetime:
 \be S=\int d^5x\sqrt{-g}(R+\frac{12}{L^2}-\frac{1}{4}F_{\mu\nu}F^{\mu\nu}-|\nabla_{\mu}\psi-iqA_{\mu}\psi|^2-m^2|\psi|^2),\ee
where $L$ is the radius of AdS spacetime. When the Maxwell field and
scalar field are absent, the above action admits the AdS soliton
solution~\cite{Horowitz:1998ha} :
 \be \label{metric} ds^2=L^2\frac{dr^2}{f(r)}+r^2(-dt^2+d\rho^2+\rho^2d\theta^2)+f(r)d\chi^2.\ee
where $f(r)=r^2-r_0^4/r^2$.\footnote{ We will work in polar
coordinates $dx^2+dy^2=d\rho^2+\rho^2d\theta^2$ in this paper.} The
asymptotical geometry approaches to $R^{1,2}\times S^1$  near the
boundary. Moreover, the Scherk-Schwarz compactification
$\chi\sim\chi+\pi L/r_0$ is needed to obtain a smooth geometry. This
gives a dual picture of a three-dimensional field theory with a mass
gap, which resembles an insulator in the condensed matter physics.
The geometry in $(r,\chi)$ directions just looks like a cigar whose
tip is at $r=r_0$. The temperature in this background is zero.

The equations of motion (EoMs) of matter fields are
 \be\label{eompsi}
 &&(\nabla_\mu-iqA_\mu)(\nabla^\mu-iqA^\mu)\psi-m^2\psi=0, \\
 \label{eomA} &&\nabla_\nu F^{\nu\mu}=iq [\bar{\psi }(\nabla^\mu-iqA^\mu)\psi-\psi(\nabla^\mu+iqA^\mu)\bar{\psi}].\ee
The boundary conditions for the matter fields near the infinity
$r\rha\infty$ of the AdS soliton are
 \be\label{bcpsi}\psi&=&\psi^{(1)}r^{-2+\sqrt{4+m^2}}+\psi^{(2)}r^{-2-\sqrt{4+m^2}}+\ldots,\\
 \label{bcA} A_t&=&\mu-\frac{\rho_e}{r^2}+\ldots,\ee
where $\psi^{(i)}=\langle \hat{O}_{(i)}\rangle,~i=1,2$, and $\hat{O}_{(i)}$ are the
corresponding dual operators of $\psi^{(i)}$
in the boundary field theory. The conformal dimensions of the
operators are $\Dl_{\pm}=2\pm\sqrt{4+m^2}$. $\mu$ and $\rho_e$
 are the corresponding chemical potential and charge density in the
 boundary field theory.

Following Ref.\cite{Nishioka:2009zj}, we  work in the probe limit,
where the Maxwell field and scalar field do not back react on the
background metric. When there is no condensate, {\it i.e.} $\psi=0$,
EoMs have the solution \be\label{Asolution} A=\mu
dt+\frac{1}{2}B\rho^2d\theta.\ee In addition to a constant chemical
potential $\mu$, adding a constant magnetic field $B$ to the Maxwell
field is also consistent in the probe limit.

As the scalar field $\psi$ does not vanish, we must solve the
coupled EoMs. However, near the critical point of phase transition
the scalar field is nearly zero, we can treat $\psi$ as a probe into
this background which is a neutral AdS soliton with a constant
electric potential and magnetic field. We are interested in
axisymmetric solutions in which all fields are independent of
$\theta$. We consider an Ansatz of the form
$\psi=F(t,r)H(\chi)U(\rho)$. Substituting $\psi$ into
Eq.\eqref{eompsi} and making the separation of the variables, we can
obtain
 \be &&\frac{\pa^2F(t,r)}{\pa r^2}+(\frac3r+\frac{\pa_rf}{f})\frac{\pa F(t,r)}{\pa
 r}-\frac{L^2}{fr^2}\frac{\pa^2F(t,r)}{\pa t^2}+\frac{2iq\mu
 L^2}{fr^2}\frac{\pa F(t,r)}{\pa
 t}\nno\\&&+\frac{L^2}{fr^2}(q^2\mu^2-m^2r^2-\frac{\la^2r^2}{f}-k^2)F(t,r)=0,\ee
where $\la^2$ and $k^2$ are the eigenvalues of the following
equations, respectively:
 \be \frac{\d^2H(\chi)}{d\chi^2}=-\la^2H(\chi),\\
 \frac{1}{\rho} \frac{d}{d\rho}[\rho \frac{d}{d\rho}U(\rho)]-\frac{1}{4}q^2 B^2\rho^2 U(\rho)=-k^2 U(\rho),\ee
where $\la=2r_0l/L$, $l\in \mathbb{Z}$ owing to the periodicity of
$H(\chi)=H(\chi+\pi L/r_0)$. $U(\rho)$ solves the equation for a
two-dimensional harmonic oscillator with frequency determined by $B$
and $k^2=n|qB|$, $n\in \mathbb{Z^+}$. We expect that the lowest mode
$l=0, n=1$ will be the first to condense and result in the most
stable solution after condensing. We can also set $L=1$ and $r_0=1$
without loss of generality. We finally obtain the equation of motion
of $F(t,r)$: \be\label{Feom} &&\frac{\pa^2F(t,r)}{\pa
r^2}+(\frac3r+\frac{\pa_rf}{f})\frac{\pa F(t,r)}{\pa
r}-\frac{L^2}{fr^2}\frac{\pa^2F(t,r)}{\pa t^2}+\frac{2iq\mu
L^2}{fr^2}\frac{\pa F(t,r)}{\pa
t}\nno\\&&+\frac{L^2}{fr^2}(q^2\mu^2-|qB|-m^2r^2)F(t,r)=0.\ee
 Note
that in this case, $U(\rho)=\exp(\frac{-|qB|\rho^2}{4})$, so for any
finite magnetic field, the superconducting condensate will be
localized to a finite circular region. As the magnetic field becomes
smaller, the region grows until it occupies the whole plane, which
can be seen in the profile of $U(\rho)$ by setting $B\rightarrow0$.
We can also see similar phenomena in other gravity
background~\cite{Hartnoll:2008kx,Albash:2008eh}.

\section{Critical behavior via Quasinormal Modes}
\label{sect:QNMs}

To reveal the stability of a spacetime background, an effective
method is to analyze the QNMs of the
perturbations in the fixed background (for reviews, see
Refs.\cite{Kokkotas:1999bd, Berti:2009kk,Konoplya:2011qq}. The temporal
part of the QNMs behave like $e^{-i\om t}$. Therefore, if the
imaginary part of the QNMs is negative, the mode will decay in time
and the perturbation will ultimately fade away, indicating that the
background is stable against this perturbation. In sharp contrast,
if the imaginary part is positive, the background is unstable. The
critical case is that if the perturbation has a marginally stable
mode, {\it i.e.} $\om=0$, one always expects that this is a signal
of instability, where a phase transition may
occur~\cite{Gubser:2008px}.

In order to study the phase transitions in the background, we
further define $F(t,r)=e^{-i\om
 t}R(r)$, and Eq.\eqref{Feom} becomes
 \be\label{Reom}
 R~''(r)+(\frac{f'}{f}+\frac3r)R~'(r)+\frac{1}{fr^2}[(\om+q\mu)^2-|qB|-m^2r^2]R(r)=0,\ee
where a prime denotes the derivative with respect to $r$.

Taking advantage of the Horowitz and Hubeny's method
\cite{Horowitz:1999jd} to study these QNMs, we find it is convenient
to work in the $z$-coordinate where $z=1/r$. In this new coordinate, the
infinite boundary is now at $z=0$, while the tip is at
$z=z_0=1/r_0=1$. Equation (\ref{Reom}) becomes \be\label{maineom}
R~''(z)+[-\frac1z+\frac{f'~(z)}{f(z)}]R~'(z)+\frac{1}{z^4f(z)}[z^2(\om+q\mu)^2-z^2|qB|-m^2]R(z)=0,\ee
where we make the prime denote the derivative with respect to $z$
from now on. Following the procedure of Horowitz and Hubeny
\cite{Horowitz:1999jd}, we multiply $z^4f(z)/(z-1)$ to both sides of
the above equation, and we obtain
 \be \label{maineq}
 S(z)R~''(z)+\frac{T(z)}{z-1}R~'(z)+\frac{V(z)}{(z-1)^2}R(z)=0,\ee
 where the coefficients are given by
 \be S(z)&=&\frac{z^4f(z)}{z-1},\\
 T(z)&=&-z^3f(z)+z^4f~'(z),\\
 \label{potential}V(z)&=&\bigg[z^2(\om+q\mu)^2-z^2|qB|-m^2\bigg](z-1).\ee
$S(z), T(z)$ and $V(z)$ are all polynomials and can be expanded to a
finite order as
 \be\label{sexpand} S(z)=\sum_{j=0}^J s_j(z-1)^j,\ee
where $J$ is a finite integer. $T(z)$ and $V(z)$ can be polynomially
expanded similarly.

Because of the absence of a black hole horizon in the AdS soliton, the
boundary condition at the tip is a finite quantity. This motivates
us to find a solution like
 \be\label{expand} R(z)=\lim_{N\rightarrow\infty}\sum_{n=0}^{N}a_n(z-1)^n.\ee
Substituting Eqs.\eqref{expand} and \eqref{sexpand} into Eq.\eqref{maineq}
and comparing the coefficients of $(z-1)^n$ for the same order, we
find that
 \be\label{an} a_n&=&-\frac{1}{H_n}\sum_{k=0}^{n-1}[k(k-1)s_{n-k}+k
 t_{n-k}+v_{n-k}]a_k,\\
 \label{hn} H_n&=&n(n-1)s_0+nt_0+v_0=-4n^2.\ee
We set $a_0=1$ for simplicity because of the linearity of
Eq.\eqref{maineq}. The boundary condition for the scalar field at
$z=0$ is
  \be\label{R0} R(0)=\lim_{N\rightarrow\infty}\sum_{n=0}^{N}a_n(-1)^n=0.\ee
And the algebraic equation \eqref{R0} can solve the modes $\om$.

\begin{figure}[h]
(A.)\includegraphics[scale=0.5]{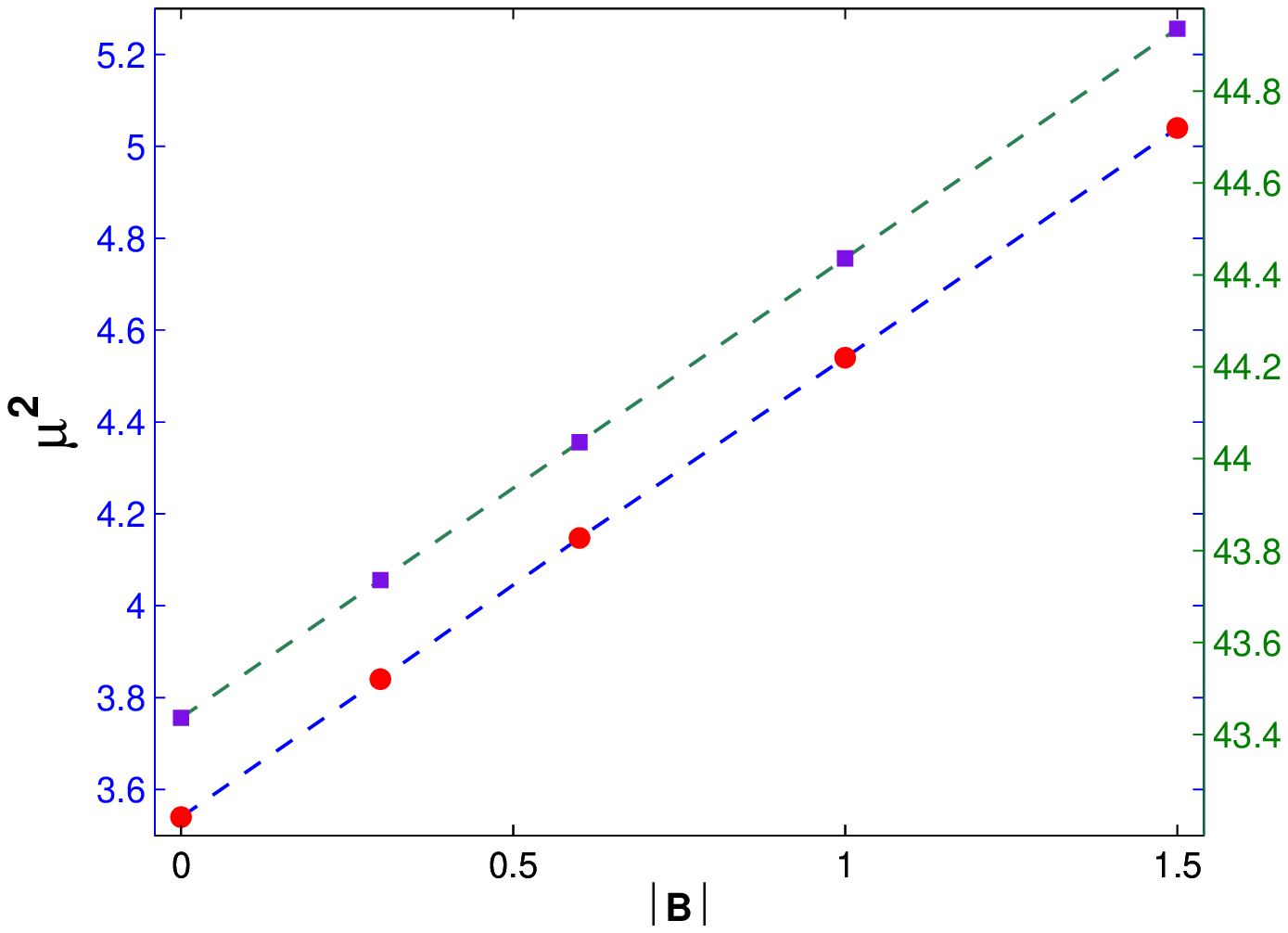}
(B.)\includegraphics[scale=0.5]{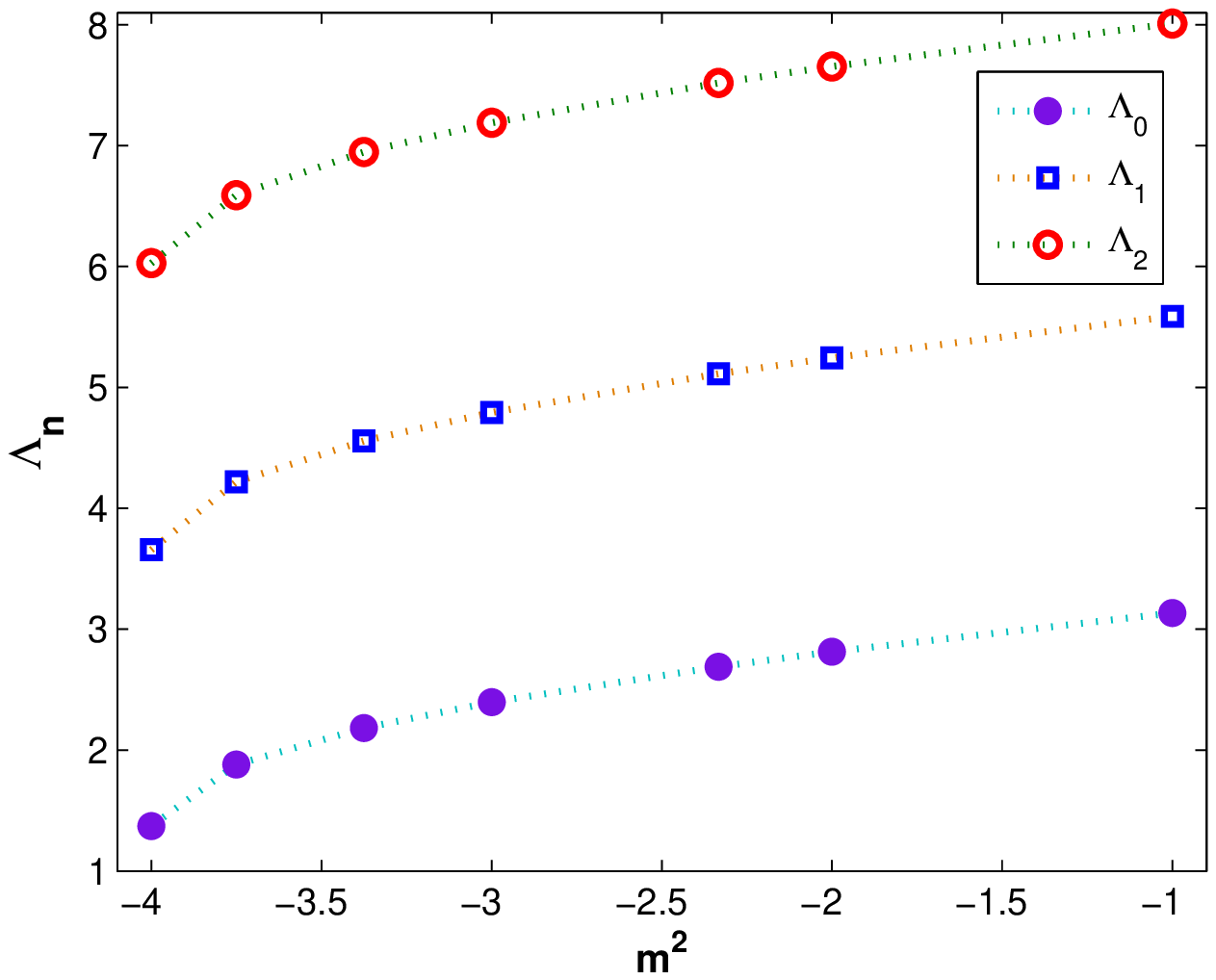} \caption{\label{qnm} (A.)
The square of the critical chemical potentials for the first (red
dots) and the third (purple quadrate points) lowest-lying modes
versus magnetic field at $m^2=-15/4$. (B.) The first three
lowest-lying $\Lambda_n$'s for the marginally stable modes versus
$m^2$ of the scalar field. The points are numerically obtained from
the QNMs.}
\end{figure}

We restrict $q=1$ just like the one in Ref.\cite{Nishioka:2009zj} in the
numerical calculations. In practice, we will expand $R(z)$ to a
large order which is $N=300$ in this paper. Because of the fact that the
real part of the frequency indicates the energy of the mode, we only
take care of the QNMs with a positive real part. From the numerical
results which are plotted in Figure.\eqref{qnm}, we find that when a
marginally stable mode arises, the square of chemical potential
$\mu^2$ and magnetic field $B$ satisfy a simple linear relation,
whose slope is particularly one, and the intercept just gives the
square of critical chemical potential in the absence of magnetic
field,  which is consistent with our previous work
\cite{Cai:2011qm}. For convenience, we define
$\mu^2-|B|\equiv\Lambda^2$ which determines the stability of the
system. The minus sign in the combination is interesting, which
means as the magnetic field grows stronger, the critical chemical
potential becomes higher. This indicates that the presence of the
magnetic field causes the transition from insulator to superconductor to be difficult.
By increasing $\Lambda$ to some critical value, the marginally stable modes will
emerge, or rather, a phase transition may occur.

\begin{table}[h]
\caption{\label{tab} The first three lowest-lying $\Lambda_n$'s for
various mass squares obtained from the calculation of the marginally
stable modes.}
\begin{center}
\begin{tabular}{cccccc}
  \hline
 $~~  ~~$ & $~~m^2=-15/4~~$ & $~~m^2=-3~~$ & $~~m^2=-7/3~~$ & $~~m^2=-2~~$ & $~~m^2=-1~~$ \\
 \hline
 $\Lambda_0$ & $1.8849$ & $2.3963$ & $2.6903$ & $2.8145$ & $3.1346$ \\
 $\Lambda_1$ & $4.2263$ & $4.7926$ & $5.1132$ & $5.2456$ & $5.5889$ \\
 $\Lambda_2$ & $6.6032$ & $7.1888$ & $7.5172$ & $7.6554$ & $8.0092$ \\
 \hline
\end{tabular}
\end{center}
\end{table}

Table \eqref{tab} shows the first three lowest-lying critical
$\Lambda_n$'s in which the index $n$ denotes the ``overtone number"
for various $m^2$'s. Marginally stable modes corresponding to higher
overtone numbers $n=1, 2\cdots$ may also appear. However, they are
unstable due to the oscillations in the $z$ direction, which can be
explicitly seen in the next section. The nodes $n=0, 1, 2$ can also
be intuitively seen in the next section. The part B of
Fig.\eqref{qnm} shows $\Lambda_n$'s of the marginally stable
modes for various squared mass of the scalar field. Turning off the
magnetic field, we recover the results previously derived in
Ref.\cite{Cai:2011qm}. Therefore, despite that we do not exactly know
the phase structures through the marginally stable modes, they can
reveal the onset of the phase transition in practice.

\section{Critical behavior via the shooting method}
\label{sect:shoot}

Shooting method is an alternative way to study the critical behavior
of the phase transition, which has been widely used in the previous
studies on holographic superconductors \cite{Hartnoll:2009sz}. In
this section, we will make use of the shooting method to study the
critical behavior in AdS soliton background, especially to plot the
profile of the scaler field. We will also compare it with the above
quasinormal modes method.

We focus on the static case in which $F$ is independent of $t$; we can
get equation of motion by simply setting $\omega=0$ in
Eq.\eqref{maineom}: \be\label{shooteom}
R~''(z)+[-\frac1z+\frac{f'~(z)}{f(z)}]R~'(z)+\frac{1}{z^4f(z)}[z^2(q^2\mu^2-|qB|)-m^2]R(z)=0.\ee
Near the boundary $R(z)$ behaves as \be R|_{z\rha0}=
\varphi^{(1)}z^{2-\sqrt{4+m^2}}+\varphi^{(2)}z^{2+\sqrt{4+m^2}}+\ldots.\ee

It is well known that when $0<\sqrt{4+m^2}<1$, the scalar field
admits two different quantizations related by a Legendre
transform~\cite{Klebanov:1999tb}. $\varphi^{(i)}$ can either be
interpreted as a source or an expectation value. In this paper, we
will only consider the case where the faster falloff is dual to the
expectation value, {\it i.e.} $\varphi^{(1)}=0$.

We now study the behavior of the solution near the tip $z=1$. Then
Eq.\eqref{shooteom} becomes  \be
R~''(z)-\frac{1}{1-z}R~'(z)+\frac{\kappa}{4(1-z)}R(z)=0,\ee where
$\kappa\equiv q^2\mu^2-|qB|-m^2\neq 0$.\footnote{Note that we
neglect the particular case $\kappa=0$. One can also consider this
case, but it does not matter.} It has a solution near $z=1$ as
 \be
R|_{z\rha1}=\alpha+\beta\log(\sqrt{|\kappa|}(1-z))+\ldots,\ee where
$\alpha$ and $\beta$ are two constants. Since we want the field to
be finite at the tip, we choose the boundary condition by setting
$\beta=0$.

To make use of the shooting method, we begin with an initial value
of $R(z)$ at the tip $z_0$ and then calculate the EoM of $R(z)$
[Eq.\eqref{shooteom}] numerically provided that the infinite boundary
condition $\varphi^{(1)}=0$ is satisfied. We also take $q=1$ for
simplicity. We can set $R(1)$ as an arbitrary constant due to the
linearity of Eq.\eqref{shooteom}. In addition, near the critical
point of the phase transition the quantity of $R(z)$ is very close
to zero, therefore, we can impose the following initial conditions
at the tip $z=1$: \be R(1)=0.001,\ \ \
R~'(1)=\frac{\kappa}{4}R(1).\ee For a given $m^2$, only for certain
values of $\mu$ and $B$ do we get to satisfy the boundary
conditions.

Note that the chemical potential $\mu$ and magnetic field $B$ appear
only as a whole $q^2\mu^2-|qB|$ in Eq.\eqref{shooteom} and
especially in $\kappa$. And inspired by the result from marginally
stable modes that only the combination $\Lambda^2\equiv\mu^2-|B|$
determines the occurrence of instability, we expect that this result
may be confirmed by our shooting method. Just as expected, we find
that only the appropriate $\Lambda^2$ can trigger the condensate.
Furthermore, the values of $\Lambda_c$ obtained from the shooting
method are perfectly consistent with the $\Lambda_n$'s in Table
\eqref{tab} derived from the ``marginally stable modes" method. So
we will not distinguish between them from now on.

\begin{figure}[h]
\includegraphics[scale=0.55]{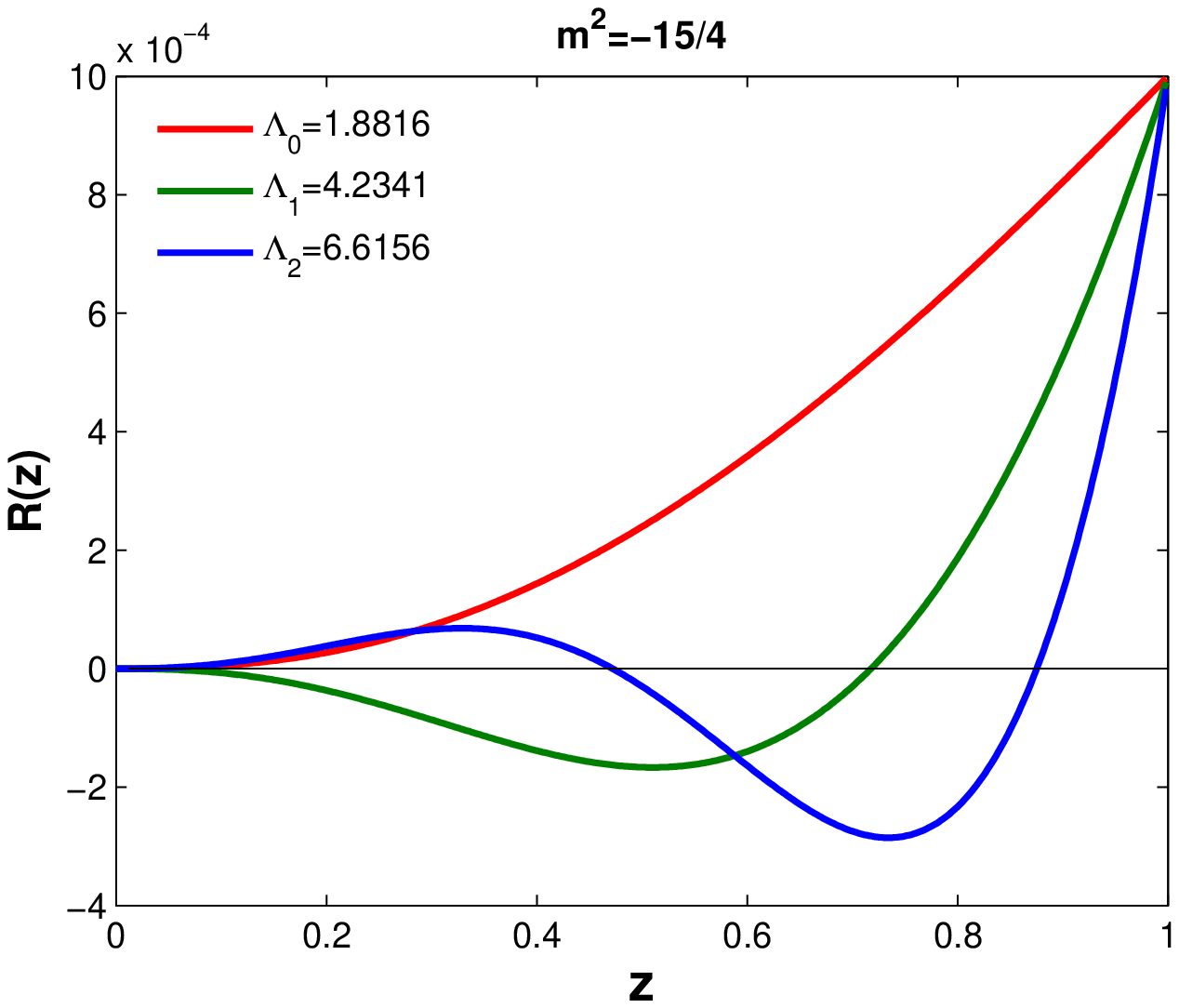}
\includegraphics[scale=0.55]{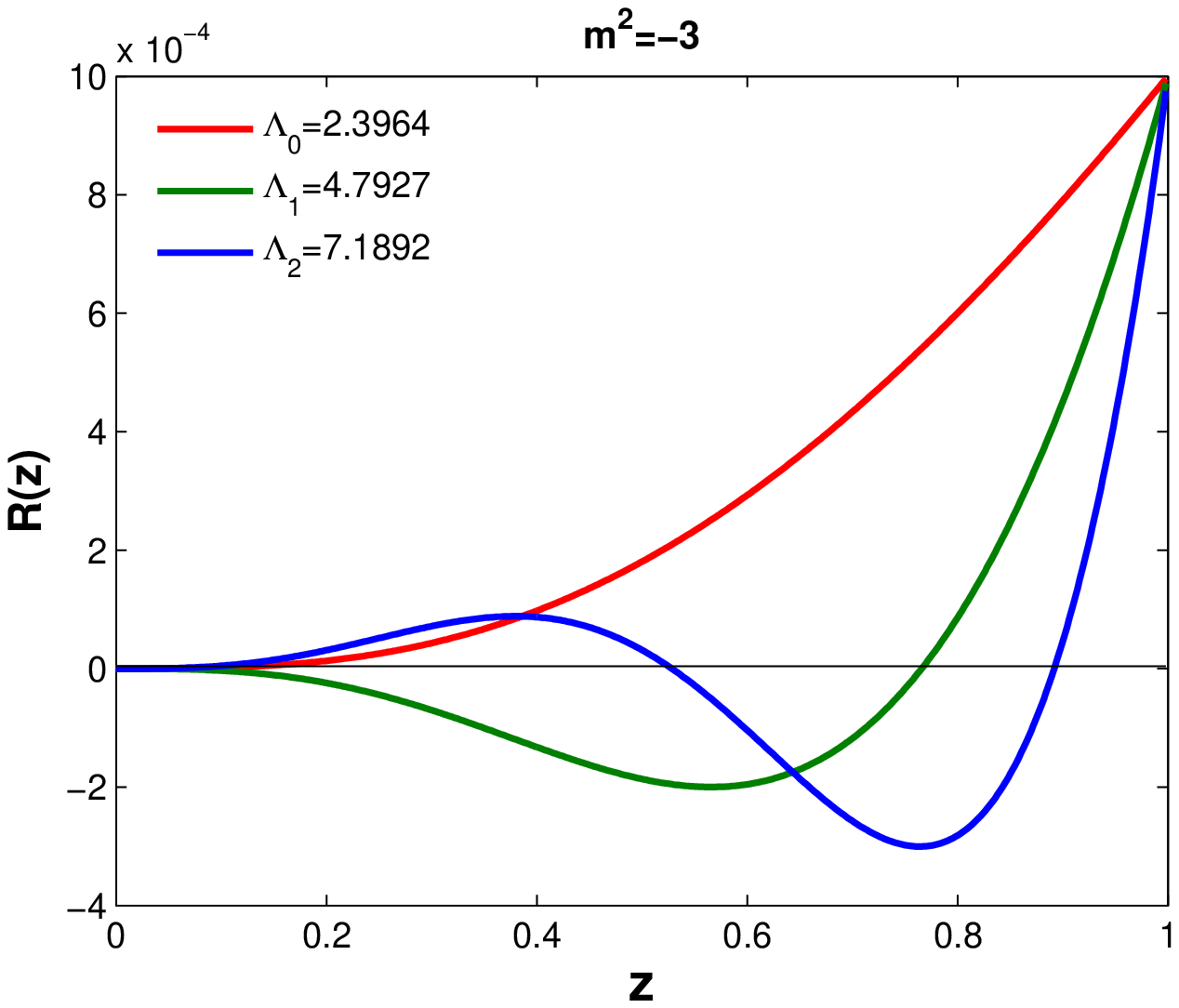}
\includegraphics[scale=0.55]{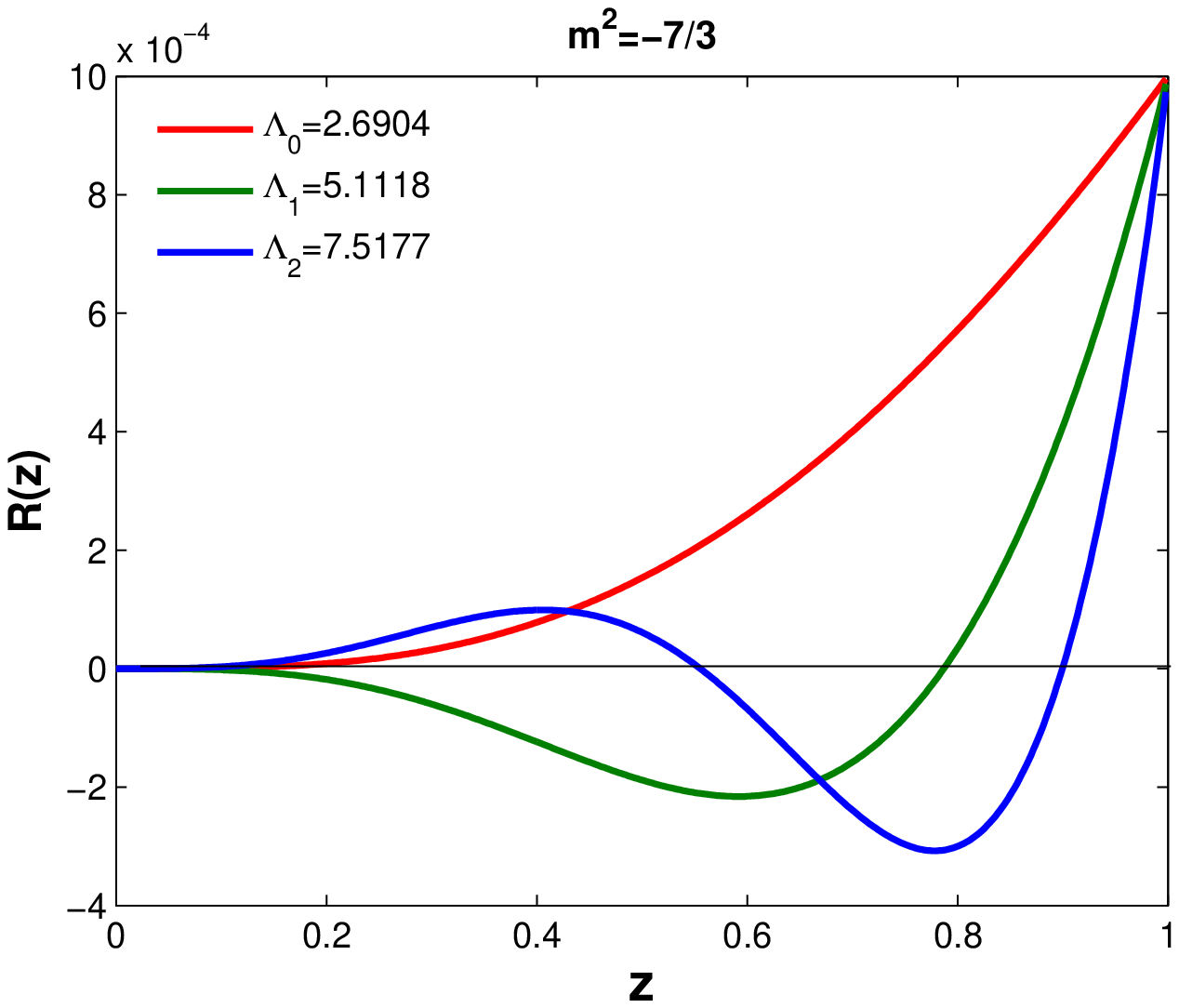}
\includegraphics[scale=0.55]{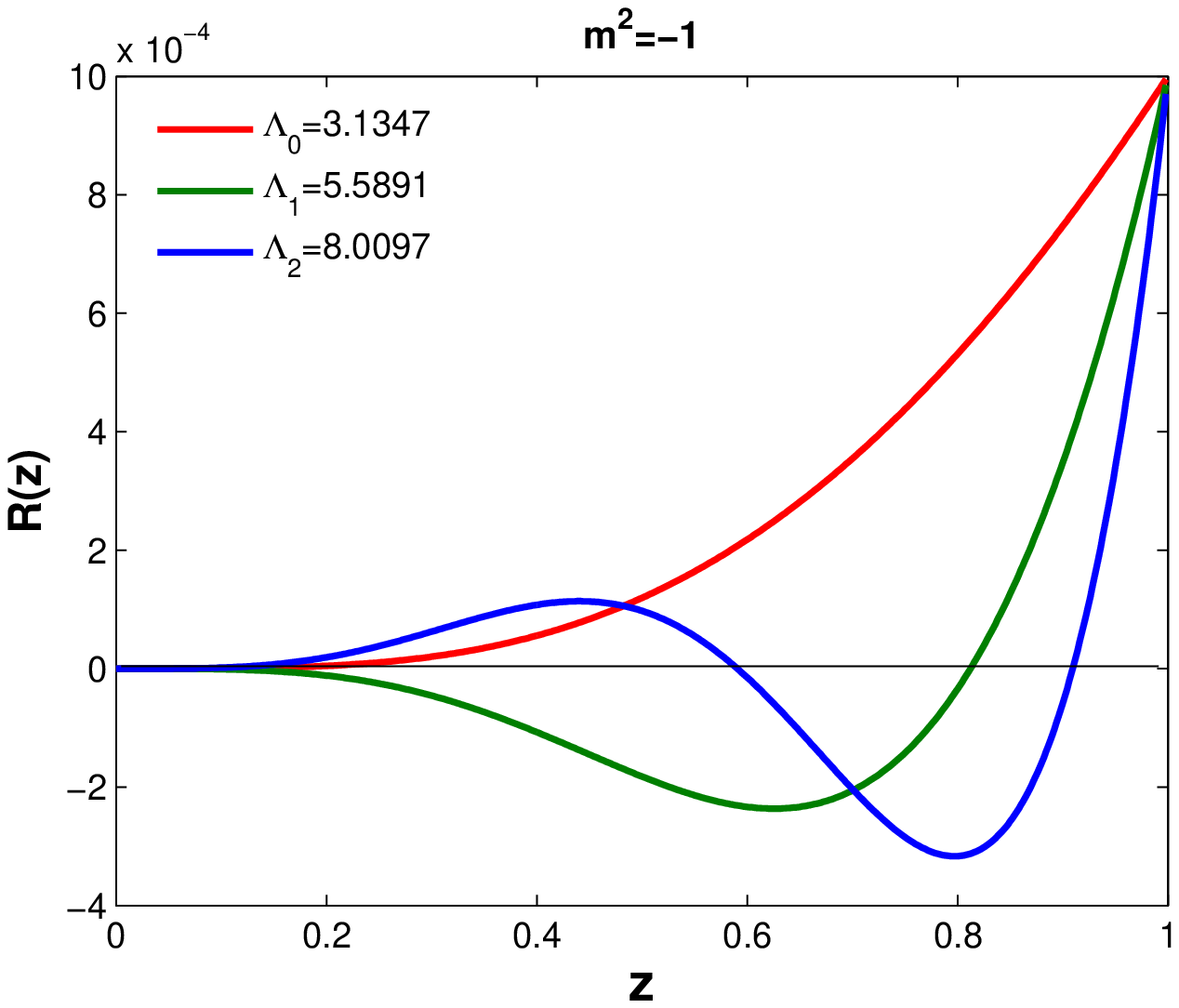}
\caption{\label{nodes} The marginally stable curves of the scalar
field corresponding to various critical $\Lambda_c$'s in the cases
of different mass square. The critical $\Lambda_c$'s for different
curves are $\Lambda_0<\Lambda_1<\Lambda_2$ in the sequence .}
\end{figure}

Figure.\eqref{nodes} exhibits the multiple marginally stable curves
of the scalar fields for various $m^2$. For example, the first three
lowest-lying modes in the plot of $m^2=-15/4$ are in the sequence
$\Lambda_0<\Lambda_1<\Lambda_2$. The red line is dual to the minimal
value of $\Lambda_c$, which has no intersecting points with the
$R(z)=0$ axis at nonvanishing $z$. Therefore, we consider the mode
corresponding to $\Lambda_c\approx1.8816$ as a mode of node $n=0$.
Further, we regard the green line dual to $\Lambda_c\approx4.2341$
and blue line related to $\Lambda_c\approx6.6156$  as modes with
nodes $n=1$ and $n=2$, respectively. However, the
 green and blue lines are thought to be unstable due to the fact
that the radial oscillations in $z$-direction of $R(z)$ will cost
energy \cite{Gubser:2008wv}. In addition, the above discussions also
hold for other diagrams in Fig.\eqref{nodes}. It is interesting
to note that the marginally stable curves corresponding to different
$\Lambda_c$'s are just the division of the curves of $R(z)$ with
different nodes.

\section{Critical behavior via the Sturm-Liouville method}
\label{sect:sturm}

From the above numerical analysis, we can see that when the
combination of chemical potential $\mu$ and magnetic field $B$, which
is $\Lambda^2\equiv\mu^2-|B|$, exceeds a critical value $\Lambda_0^2$
for given mass, the condensations of the operators will turn out.
This can be regarded as a superconductor (superfluid) phase.
However, when less than $\Lambda_0$, the scalar field is vanishing
and this can be interpreted as the insulator phase because this
system has a mass gap, which is due to the confinement in the
(2+1)-dimensional gauge theory via the Scherk-Schwarz
compactificaiton. Therefore, the critical parameters satisfied,
$\Lambda_0=\mu^2-|B|$, are the turning points of this holographic
insulator/superconductor phase transition.

In this section, using the variational method for the
Sturm-Liouville eigenvalue problem \cite{Siopsis:2010uq}, we
analytically study the phase transition just following the procedure
in Ref.\cite{Cai:2011ky}. Especially, we look forward to giving an
explicit demonstration to our previous numerical results and trying
to find an approximate function to relate parameters
$\{q,\mu,B,m^2\}$ at the critical phase transition point. We start
from Eq.\eqref{shooteom} and also focus on the case
$\varphi^{(1)}=0$. The operator $\hat{O}_{(2)}$ is normalizable when
$m^2>m_{\rm BF}^2=-4$, where $m_{\rm BF}^2$ is the
Breitenlohner-Freedman bound of the mass square of scalar field
in the AdS spacetime.

\begin{figure}[h]
(A.)\includegraphics[scale=0.5]{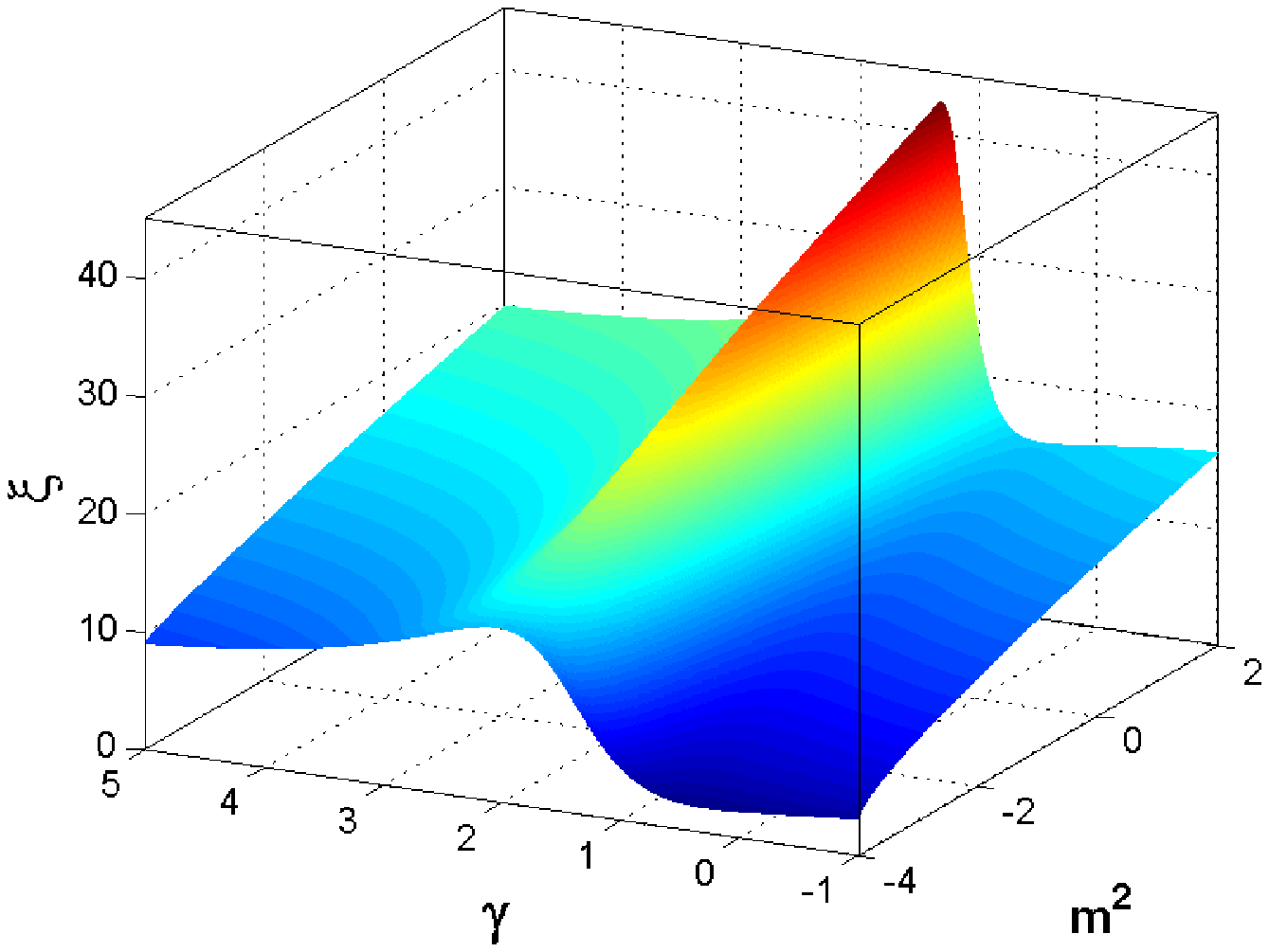}
(B.)\includegraphics[scale=0.5]{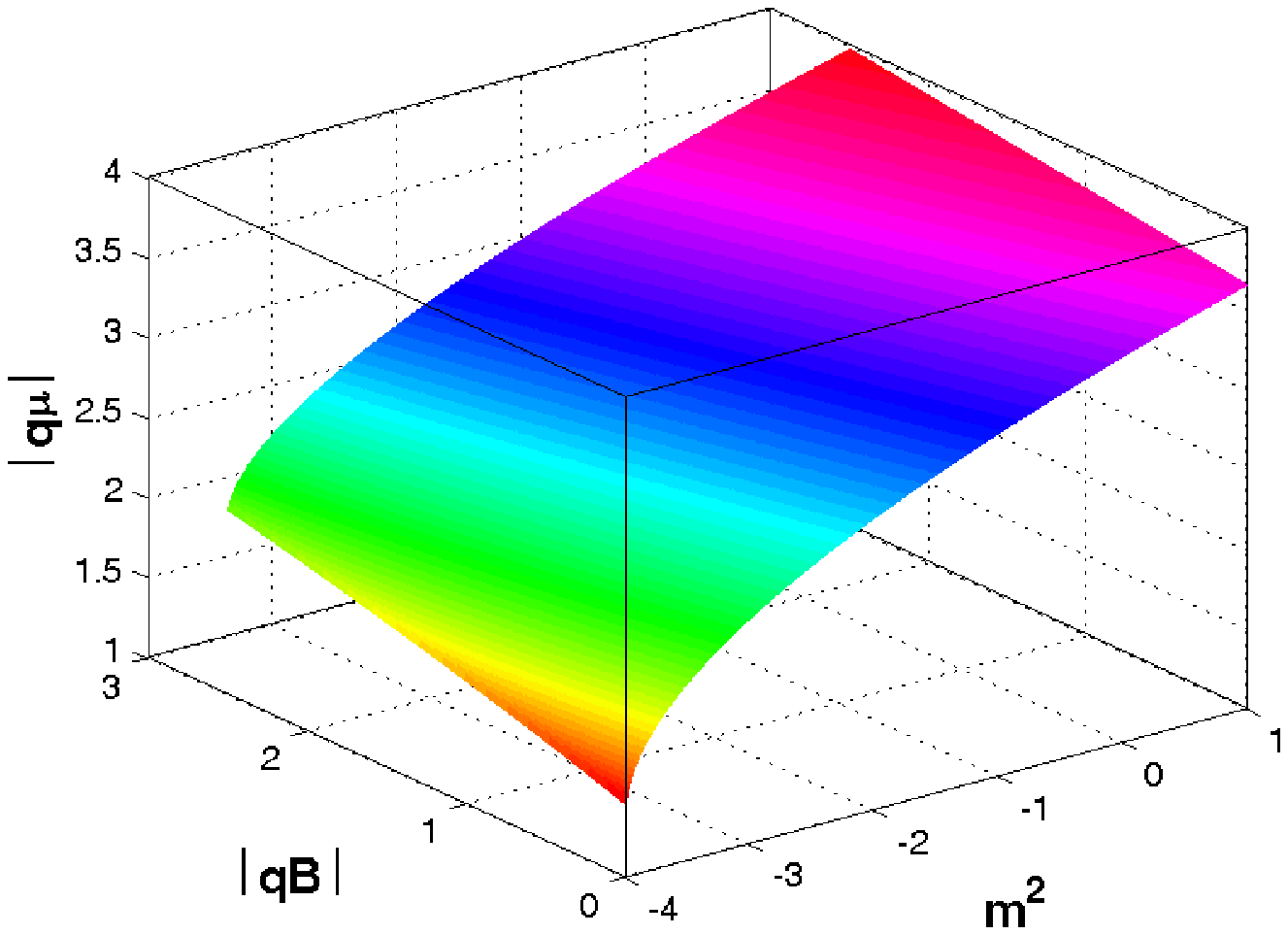}
(C.)\includegraphics[scale=0.5]{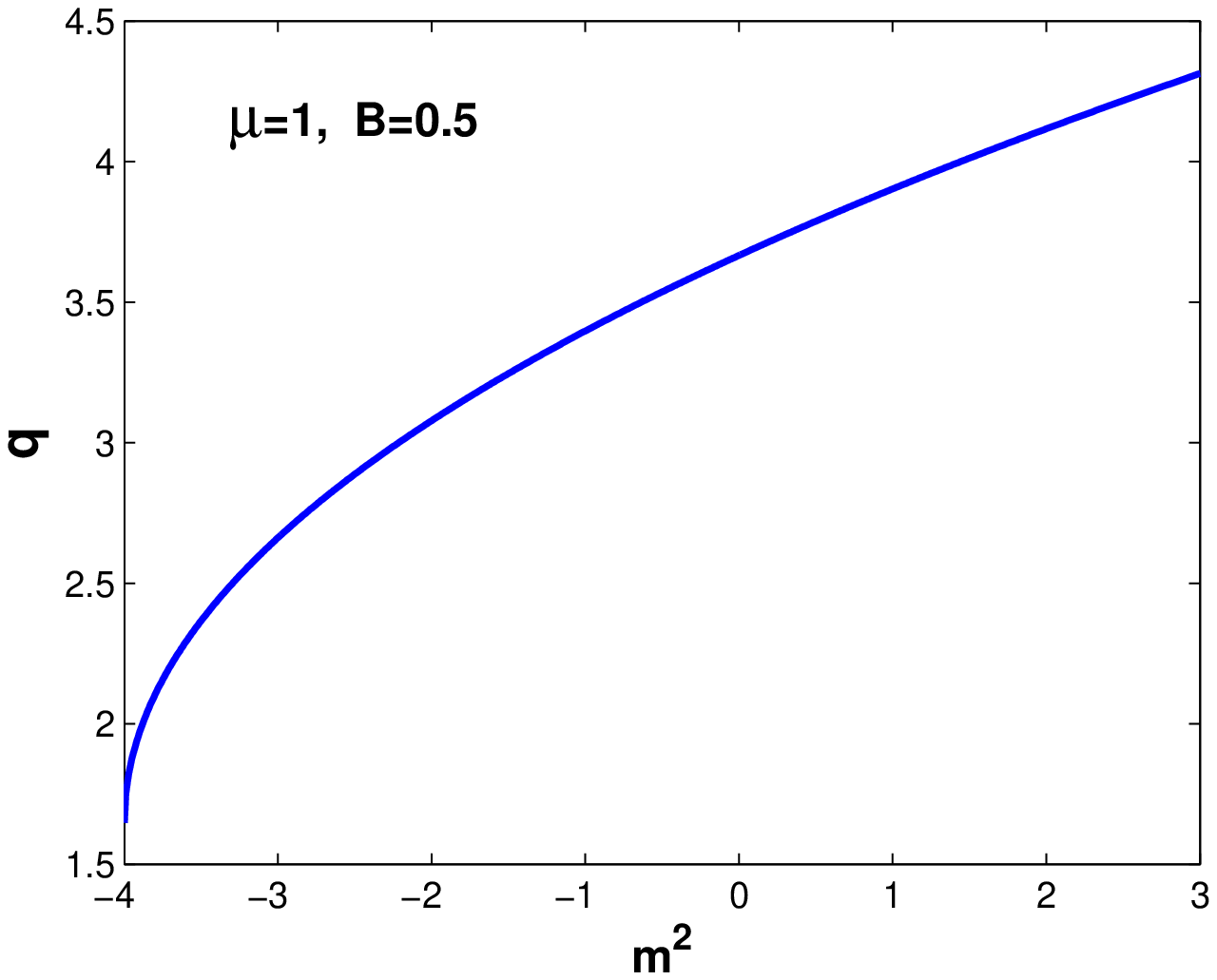}
(D.)\includegraphics[scale=0.5]{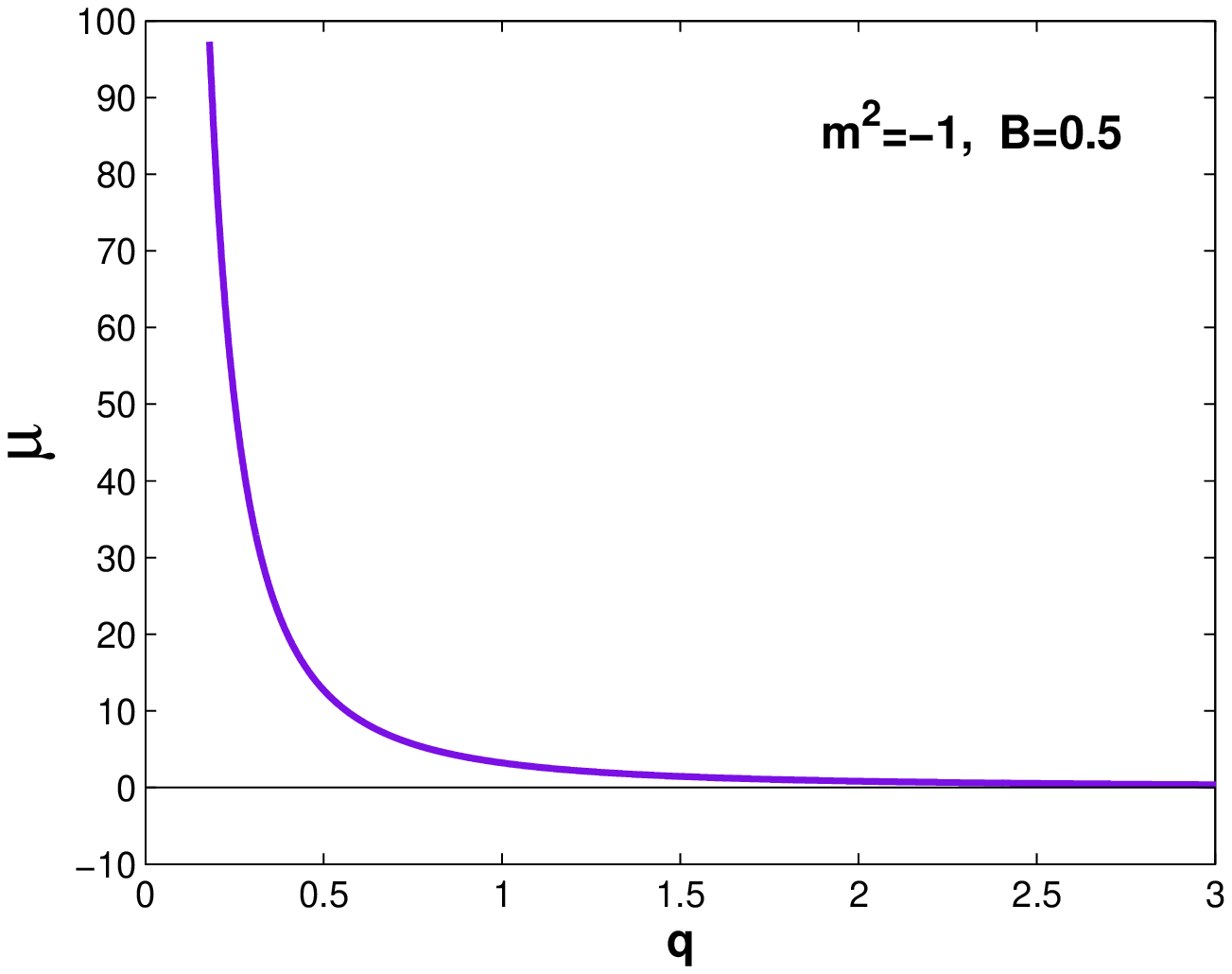}
 \caption{\label{musquare}(A.) $\xi$ as a function of
 $\gamma$ and $m^2$. (B.) The critical chemical potenial
 $|q\mu|$ versus $|qB|$ and $m^2$. (C.) The
 charge $q$ versus mass square $m^2$ at fixed chemical potential
 $\mu=1$ and magnetic field $B=0.5$. (D.) The
  critical chemical potential $\mu$ versus the charge $q$ at fixed
 mass square
 $m^2=-1$ and magnetic field $B=0.5$.}
\end{figure}

Following the steps in Ref.\cite{Cai:2011ky}, we introduce a trial
function $\Gamma(z)$ into $R(z)$ near $z=0$ as
  \be
R|_{z\rha0}\simeq\langle \hat{O}_{(2)}\rangle
z^{2+\sqrt{4+m^2}}\Gamma(z).\ee
 The boundary conditions for $\Gamma(z)$ are $\Gamma(0)=1$ and $\Gamma'(0)=0$.
 It is easy to obtain the EoM of $\Gamma(z)$ as
 \be &&
 \Gamma~''(z)+\frac{z^4(5+2\sqrt{4+m^2})-1-2\sqrt{4+m^2}}{z(z^4-1)}\Gamma~'(z)+
 \frac{|qB|-q^2\mu^2}{z^4-1}\Gamma(z)
 \nno\\&& +\frac{z^2(8+m^2+4\sqrt{4+m^2})}{z^4-1}\Gamma(z)=0.\ee
The EoM of $\Gamma(z)$ can be rewritten as \be &&
\frac{d}{dz}[\underbrace{z^{1+2\sqrt{4+m^2}}(z^4-1)}_K\Gamma~'(z)]+\underbrace{z^{3+2\sqrt{4+m^2}}(8+m^2+4\sqrt{4+m^2})}_{-P}\Gamma(z)
\nno\\&&
\underbrace{-z^{1+2\sqrt{4+m^2}}}_Q(q^2\mu^2-|qB|)\Gamma(z)=0.\ee
 The minimum eigenvalue of $q^2 \mu^2-|qB|$ can be obtained by varying
 the following functional:
\be
q^2\mu^2-|qB|&=&\frac{\int_0^1dz(K\Gamma'^2+P\Gamma^2)}{\int_0^1dz~Q\Gamma^2}\equiv\xi(\gamma,m^2).\ee
In order to estimate it, we try to set $\Gamma(z)=1-\gamma z^2$.
Thus, we obtain a particular function $q^2\mu^2-|qB|=\xi(\gamma,m^2)$,
from which we can get the minimal value of $q^2\mu^2-|qB|$ for given
mass. It will be subtle if the function $\xi(\gamma,m^2)$ has more
than one minimum. Fortunately, we can see from part A of
Fig.\eqref{musquare} that there indeed exists only one minimum
for given mass square. We denote the minimum for given  mass square
as $\Lambda[m^2]$.

\begin{table}[h]
\caption{\label{tab2} The critical parameters $q^2\mu^2-|qB|$ for
various mass squares obtained from the QNMs, shooting method, and
S-L method.}
\begin{center}
\begin{tabular}{cccccc}
  \hline
 $~~  ~~$ & $~~m^2=-15/4~~$ & $~~m^2=-3~~$ & $~~m^2=-7/3~~$ & $~~m^2=-1~~$ \\
 \hline
 ${\rm QNMs}$      & $1.8849$ & $2.3963$ & $2.6903$ & $3.1346$ \\
 ${\rm Shooting}$  & $1.8816$ & $2.3964$ & $2.6904$ & $3.1347$ \\
 ${\rm S-L}$       & $1.8904$ & $2.3986$ & $2.6928$ & $3.1373$ \\
 \hline
\end{tabular}
\end{center}
\end{table}
In Table \eqref{tab2}, we list the critical parameters $\Lambda_0$
for various mass squares obtained from the three methods, {\it i.e.}
the calculation of the marginally stable modes, the shooting method,
and the Sturm-Liouville (S-L) method. The analytical results are in
good agreement with numerical values. So we can trust the analytical
treatment, especially the choice of $\Gamma(z)=1-\gamma z^2$.

We finally obtain the simple relation for parameters $q,\mu,B,m^2$
at the critical phase transition point as \be
q^2\mu^2-|qB|=\Lambda[m^2].\ee This simple relation can explain the
main results we obtained. We plot the function
$|q\mu|=|q\mu|(|qB|,m^2)$ in part B of Fig.\eqref{musquare}.
The relation can be regarded as a function of three arbitrary
independent variables in $\{q,\mu,B,m^2\}$. We extract the relation
between the charge $q$ and mass square $m^2$ for the systems whose
phase transitions take place at fixed chemical potential $\mu$ and
magnetic field $B$. Part C of Fig.\eqref{musquare} shows
schematically that when the mass square of the scalar field grows,
the charge must grow too. Similarly, it can be seen from part D
of Fig.\eqref{musquare} that the critical chemical potentials are
sufficiently depressed by the charge of the scalar field for fixed
mass and magnetic field.

\section{Discussions and Conclusions}
\label{sect:con}

In the probe limit, we studied the holographic
insulator/superconductor transition in an external constant magnetic
field both numerically and analytically. When the magnetic field is
absent, the condensate fills the plane homogeneously. On the other
hand, noting that the profile of scalar field
$\psi(z,\rho)=R(z)\exp(\frac{-|qB|\rho^2}{4})$ where the profile of
$R(z)$ in superconducting phase is shown in Fig.\eqref{nodes},
we see that for any nonvanishing magnetic field, the superconducting condensate
 will be localized to a
finite circular region. As the magnetic field becomes smaller the
region grows until it occupies the whole plane, which can be seen in
profile of $U(\rho)$ by setting $B\rightarrow0$.

Despite that we do not exactly know the phase structure by
investigating the marginally stable modes of the scalar field, they
can actually indicate the onset of the phase transition, which means
that the neutral AdS soliton  will become unstable to develop
charged scalar hairs in this AdS soliton background when the
parameters are beyond the critical values. Our results show that
only the parameter $\Lambda^2=q^2\mu^2-|qB| $ triggers the
instability. Making use of the shooting method to numerically plot
the behavior of the scalar field in the radial direction, one can
intuitively see the nodes of the marginally stable modes.
Furthermore, taking advantage of analytical approach we directly
obtained a simple relation which constrains parameters
$\{q,\mu,B,m^2\}$ at the critical point, {\it i.e.}
$q^2\mu^2-|qB|=\Lambda[m^2]$. This relation is the main result of
our paper, and it can be used to explain many properties of the
model.

Note that in this paper, we have limited ourselves to the choice of
$\psi^{(1)}=0$ and $\lambda=0$, and $n=1$. It is interesting to study the
cases with vanishing $\psi^{(2)}$ and other $(\lambda, n)$'s. Of
course, it is also quite significant to investigate the back
reaction of matter fields and to draw a full phase diagram of the
holographic insultor/superconductor transition. These discussions in
the paper are desired to extend to the background of modified
gravity, such as including Chern-Simons term and $R^2$
term\cite{Li:2011xj,Takeuchi:2011uk}. We expect to report on further
studies on these and relevant issues.

\section*{Acknowledgements}
We would like to thank Zhang-Yu Nie for his indispensable help. This
work was supported in part by the National Natural Science
Foundation of China (No. 10821504, No. 10975168 and No.11035008),
the Ministry of Science and Technology of China under Grant No.
2010CB833004, and a grant from the Chinese Academy of Sciences.

%\appendix


\begin{thebibliography}{99}
\baselineskip 12pt

%\cite{Maldacena:1997re}
\bibitem{Maldacena:1997re}
  J.~M.~Maldacena,
  ``The large N limit of superconformal field theories and supergravity,''
  Adv.\ Theor.\ Math.\ Phys.\  {\bf 2}, 231 (1998)
  [Int.\ J.\ Theor.\ Phys.\  {\bf 38}, 1113 (1999)]
  [arXiv:hep-th/9711200].
  %%CITATION = IJTPB,38,1113;%%

%\cite{Gubser:2008px}
\bibitem{Gubser:2008px}
  S.~S.~Gubser,
  ``Breaking an Abelian gauge symmetry near a black hole horizon,''
  Phys.\ Rev.\  D {\bf 78}, 065034 (2008)
  [arXiv:0801.2977 [hep-th]].
  %%CITATION = PHRVA,D78,065034;%%

%\cite{Hartnoll:2008vx}
\bibitem{Hartnoll:2008vx}
  S.~A.~Hartnoll, C.~P.~Herzog and G.~T.~Horowitz,
  ``Building a Holographic Superconductor,''
  Phys.\ Rev.\ Lett.\  {\bf 101}, 031601 (2008)
  [arXiv:0803.3295 [hep-th]].
  %%CITATION = PRLTA,101,031601;%%

%\cite{Hartnoll:2009sz}
\bibitem{Hartnoll:2009sz}
  S.~A.~Hartnoll,
  ``Lectures on holographic methods for condensed matter physics,''
  Class.\ Quant.\ Grav.\  {\bf 26}, 224002 (2009)
  [arXiv:0903.3246 [hep-th]].
  %%CITATION = CQGRD,26,224002;%%

%\cite{Hartnoll:2008kx}
\bibitem{Hartnoll:2008kx}
  S.~A.~Hartnoll, C.~P.~Herzog, G.~T.~Horowitz,
  ``Holographic Superconductors,''
  JHEP {\bf 0812}, 015 (2008).
  [arXiv:0810.1563 [hep-th]].


%\cite{Nishioka:2009zj}
\bibitem{Nishioka:2009zj}
  T.~Nishioka, S.~Ryu, T.~Takayanagi,
  ``Holographic Superconductor/Insulator Transition at Zero Temperature,''
  JHEP {\bf 1003}, 131 (2010).
  [arXiv:0911.0962 [hep-th]].

%\cite{Horowitz:1998ha}
\bibitem{Horowitz:1998ha}
  G.~T.~Horowitz, R.~C.~Myers,
  ``The AdS / CFT correspondence and a new positive energy conjecture for general relativity,''
  Phys.\ Rev.\  {\bf D59}, 026005 (1998).
  [hep-th/9808079].


%\cite{Witten:1998zw}
\bibitem{Witten:1998zw}
  E.~Witten,
  ``Anti-de Sitter space, thermal phase transition, and confinement in gauge theories,''
  Adv.\ Theor.\ Math.\ Phys.\  {\bf 2}, 505-532 (1998).
  [hep-th/9803131].

  %\cite{Horowitz:2010jq}
\bibitem{Horowitz:2010jq}
  G.~T.~Horowitz, B.~Way,
  ``Complete Phase Diagrams for a Holographic Superconductor/Insulator System,''
  JHEP {\bf 1011}, 011 (2010).
  [arXiv:1007.3714 [hep-th]].

  %\cite{Akhavan:2010bf}
\bibitem{Akhavan:2010bf}
  A.~Akhavan, M.~Alishahiha,
  ``P-Wave Holographic Insulator/Superconductor Phase Transition,''
  [arXiv:1011.6158 [hep-th]].

%\cite{Basu:2011yg}
\bibitem{Basu:2011yg}
  P.~Basu, F.~Nogueira, M.~Rozali, J.~B.~Stang and M.~Van Raamsdonk,
  ``Towards A Holographic Model of Color Superconductivity,''
  arXiv:1101.4042 [hep-th].
  %%CITATION = ARXIV:1101.4042;%%

  %\cite{Brihaye:2011vk}
\bibitem{Brihaye:2011vk}
  Y.~Brihaye and B.~Hartmann,
  ``Holographic superfluid/fluid/insulator phase transitions in 2+1
  dimensions,''
  arXiv:1101.5708 [hep-th].
  %%CITATION = ARXIV:1101.5708;%%

%\cite{Cai:2011ky}
\bibitem{Cai:2011ky}
  R.~G.~Cai, H.~F.~Li and H.~Q.~Zhang,
  ``Analytical Studies on Holographic Insulator/Superconductor Phase
  Transitions,''
  arXiv:1103.5568 [hep-th].
  %%CITATION = ARXIV:1103.5568;%%


%\cite{Peng:2011gh}
\bibitem{Peng:2011gh}
  Y.~Peng, Q.~Pan, B.~Wang,
  ``Various types of phase transitions in the AdS soliton background,''
  Phys.\ Lett.\  {\bf B699}, 383-387 (2011).
  [arXiv:1104.2478 [hep-th]].

%\cite{Pan:2011ah}
\bibitem{Pan:2011ah}
  Q.~Pan, J.~Jing, B.~Wang,
  ``Analytical investigation of the phase transition between holographic insulator and superconductor in Gauss-Bonnet gravity,''
  [arXiv:1105.6153 [gr-qc]].


%\cite{Nakano:2008xc}
\bibitem{Nakano:2008xc}
  E.~Nakano, W.~-Y.~Wen,
  ``Critical magnetic field in a holographic superconductor,''
  Phys.\ Rev.\  {\bf D78}, 046004 (2008).
  [arXiv:0804.3180 [hep-th]].

%\cite{Albash:2008eh}
\bibitem{Albash:2008eh}
  T.~Albash, C.~V.~Johnson,
  ``A Holographic Superconductor in an External Magnetic Field,''
  JHEP {\bf 0809}, 121 (2008).
  [arXiv:0804.3466 [hep-th]].

%\cite{Wen:2008pb}
\bibitem{Wen:2008pb}
  W.~Y.~Wen,
  ``Inhomogeneous magnetic field in AdS/CFT superconductor,''
  arXiv:0805.1550 [hep-th].
  %%CITATION = ARXIV:0805.1550;%%

%\cite{Maeda:2008ir}
\bibitem{Maeda:2008ir}
  K.~Maeda, T.~Okamura,
  ``Characteristic length of an AdS/CFT superconductor,''
  Phys.\ Rev.\  {\bf D78}, 106006 (2008).
  [arXiv:0809.3079 [hep-th]].

%\cite{Ge:2010aa}
\bibitem{Ge:2010aa}
  X.~-H.~Ge, B.~Wang, S.~-F.~Wu, G.~-H.~Yang,
  ``Analytical study on holographic superconductors in external magnetic field,''
  JHEP {\bf 1008}, 108 (2010).
  [arXiv:1002.4901 [hep-th]].

%\cite{arXiv:0903.1864}
\bibitem{arXiv:0903.1864}
        M.~Ammon, J.~Erdmenger, M.~Kaminski and P.~Kerner,
        %``Flavor Superconductivity from Gauge/Gravity Duality,''
JHEP\ {\bf 0910}, 067  (2009)
[arXiv:0903.1864 [hep-th]].
%%CITATION = JHEPA,0910,067;%%

%\cite{Cai:2011qm}
\bibitem{Cai:2011qm}
  R.~-G.~Cai, X.~He, H.~-F.~Li, H.~-Q.~Zhang,
  ``Phase transitions in AdS soliton spacetime through marginally stable modes,''
  Phys.\ Rev.\  {\bf D84}, 046001 (2011).
  [arXiv:1105.5000 [hep-th]].




%\cite{Horowitz:1999jd}
\bibitem{Horowitz:1999jd}
  G.~T.~Horowitz and V.~E.~Hubeny,
  ``Quasinormal modes of AdS black holes and the approach to thermal equilibrium,''
  Phys.\ Rev.\  D {\bf 62}, 024027 (2000)
  [arXiv:hep-th/9909056].
  %%CITATION = PHRVA,D62,024027;%%

%\cite{Gubser:2008wv}
\bibitem{Gubser:2008wv}
  S.~S.~Gubser and S.~S.~Pufu,
  ``The Gravity dual of a p-wave superconductor,''
  JHEP {\bf 0811}, 033 (2008)
  [arXiv:0805.2960 [hep-th]].
  %%CITATION = JHEPA,0811,033;%%

%\cite{Siopsis:2010uq}
\bibitem{Siopsis:2010uq}
  G.~Siopsis, J.~Therrien,
  ``Analytic calculation of properties of holographic superconductors,''
  JHEP {\bf 1005}, 013 (2010).
  [arXiv:1003.4275 [hep-th]].

%\cite{Kokkotas:1999bd}
\bibitem{Kokkotas:1999bd}
  K.~D.~Kokkotas and B.~G.~Schmidt,
  ``Quasinormal modes of stars and black holes,''
  Living Rev.\ Rel.\  {\bf 2}, 2 (1999)
  [arXiv:gr-qc/9909058].
  %%CITATION = 00222,2,2;%%

%\cite{Berti:2009kk}
\bibitem{Berti:2009kk}
  E.~Berti, V.~Cardoso and A.~O.~Starinets,
  ``Quasinormal modes of black holes and black branes,''
  Class.\ Quant.\ Grav.\  {\bf 26}, 163001 (2009)
  [arXiv:0905.2975 [gr-qc]].
  %%CITATION = CQGRD,26,163001;%%

%\cite{Konoplya:2011qq}
\bibitem{Konoplya:2011qq}
  R.~A.~Konoplya and A.~Zhidenko,
  ``Quasinormal modes of black holes: From astrophysics to string theory,''
  arXiv:1102.4014 [gr-qc].
  %%CITATION = ARXIV:1102.4014;%%

%\cite{Klebanov:1999tb}
\bibitem{Klebanov:1999tb}
  I.~R.~Klebanov and E.~Witten,
  ``AdS / CFT correspondence and symmetry breaking,''
  Nucl.\ Phys.\  B {\bf 556} (1999) 89
  [arXiv:hep-th/9905104].
  %%CITATION = NUPHA,B556,89;%%

%\cite{Li:2011xj}
\bibitem{Li:2011xj}
  H.~-F.~Li, R.~-G.~Cai, H.~-Q.~Zhang,
  ``Analytical Studies on Holographic Superconductors in Gauss-Bonnet Gravity,''
  JHEP {\bf 1104}, 028 (2011).
  [arXiv:1103.2833 [hep-th]].

%\cite{Takeuchi:2011uk}
\bibitem{Takeuchi:2011uk}
  S.~Takeuchi,
  ``Modulated Instability in Five-Dimensional U(1) Charged AdS Black Hole with R**2-term,''
 [arXiv:1108.2064 [hep-th]].


\end{thebibliography}
\end{document}